\documentclass{article}
\usepackage{spconf,amsmath,graphicx}
\usepackage{spconf,amsmath,graphicx}
\usepackage{amssymb}
\usepackage{multicol}
\usepackage{multirow}
\usepackage{multirow}
\usepackage{subfigure}
\usepackage{graphicx}
\usepackage{array}
\usepackage{amssymb}
\usepackage{tikz}
\usepackage{amsmath}
\usepackage{changepage}
\usepackage{tabularx}
\usepackage{color}
\usepackage{booktabs}
\usepackage{siunitx}
\usepackage{arydshln}
\usepackage{utfsym}
\usepackage{dashrule}
\usepackage{algorithm,algorithmic}

\title{FGCL: Fine-grained Contrastive Learning For Mandarin Stuttering Event Detection}
 \name{Han Jiang{\normalfont\textsuperscript{1}}$^{\ast}$, Wenyu Wang{\normalfont\textsuperscript{1}}$^{\ast}$\thanks{$^{\ast}$Equal contribution. $^{\dag}$Corresponding author.}, Yiquan Zhou{\normalfont\textsuperscript{1}}$^{\dag}$ , Hongwu Ding{\normalfont\textsuperscript{2}}, Jiacheng Xu{\normalfont\textsuperscript{2}}, Jihua Zhu{\normalfont\textsuperscript{1}}}

\address{\textsuperscript{1}School of Software Engineering, Xi’an Jiaotong University, Xi'an, China\\
        \textsuperscript{2}Happy Elements, Shanghai, China}

\begin{document}
%
\maketitle
\begin{abstract}

This paper presents the T031 team's approach to the StutteringSpeech Challenge in SLT2024. Mandarin Stuttering Event Detection (MSED) aims to detect instances of stuttering events in Mandarin speech. We propose a detailed acoustic analysis method to improve the accuracy of stutter detection by capturing subtle nuances that previous Stuttering Event Detection (SED) techniques have overlooked. To this end, we introduce the Fine-Grained Contrastive Learning (FGCL) framework for MSED. Specifically, we model the frame-level probabilities of stuttering events and introduce a mining algorithm to identify both \textit{easy} and \textit{confusing} frames. Then, we propose a stutter contrast loss to enhance the distinction between stuttered and fluent speech frames, thereby improving the discriminative capability of stuttered feature embeddings. Extensive evaluations on English and Mandarin datasets demonstrate the effectiveness of FGCL, achieving a significant increase of over 5.0\% in F1 score on Mandarin data\footnote{FGCL won 3rd place in Mandarin stuttering event detection and automatic speech recognition in SLT2024}.

\end{abstract}
\section{Introduction}
\label{sec:intro}

Speech is a tool used by humans for communication, allowing people to convey information and interact with one another. Stuttering, as a speech disorder, is typically characterized by prolonged sounds, repeated sounds, repeated characters, blockages, and the use of interjections as fillers \cite{guitar2013stuttering}. Stuttering affects normal social activities, which can lead to stress, shame, and inferiority in affected individuals, thereby causing negative emotions and harming mental health. The automated identification of stuttering behaviors, a process formally recognized as Stuttering Event Detection (SED), entails the systematic analysis of a speech sample, wherein specialized systems discern the presence or absence of disfluencies indicative of stuttering.

With the development of neural networks, deep neural network-based stuttering event detection systems have made significant progress. \cite{sheikh2021stutternet} proposed a Time Delay Neural Network (TDNN) architecture for stuttering detection. \cite{al2022stuttering} employed a 2D atrous convolutional network to extract spectral and temporal features for stuttering detection. \cite{kourkounakis2020mfcc3} introduced a disfluency detection network based on ResNet and BiLSTM. \cite{jouaiti2022mfcc4} incorporate phoneme probability and mel-frequency cepstral coefficients (MFCCs) to train a neural network for the detection and classification of stuttering. Despite the notable advancements in neural network-driven stuttering detection systems, there is still significant room for improvement. Current approaches often neglect the crucial aspect of fine-grained modeling, which is essential for capturing the subtle nuances present in stuttering speech patterns.

Recently, contrastive learning \cite{chen2020simple,he2020momentum,zhang2021cola} and self-supervised learning (SSL) \cite{baevski2020wav2vec,hsu2021hubert} have been widely applied to various tasks in the field of speech processing. Contrastive learning is proven to be effective in representation learning without ground-truth labels, and SSL models in speech have shown impressive performance in downstream tasks. \cite{mohapatra2022ssl2} utilized contextual embeddings from pretrained networks to design a disfluency classification network. \cite{al2024ssl3} introduced a method based on multi-feature fusion and attention mechanisms, leveraging various features from different pitch, time-domain, frequency-domain aspects, and SSL features. DisfluentSiam \cite{mohapatra2023ssl1} proposed a lightweight pretraining pipeline based on wav2vec2.0 \cite{baevski2020wav2vec} and Siamese networks. Inspired by the impact of contrastive learning and self-supervised learning on speech-related tasks, we are driven to extend these techniques to the SED task, aiming to elevate the state-of-the-art in this area through innovative application.

Despite the availability of extensive English-language datasets for stuttering event detection, such as SEP-28k \cite{lea2021sep} and FluencyBank \cite{ratner2018fluency}, a significant gap exists in research related to Mandarin stuttering event detection. To address this gap, the StutteringSpeech Challenge \cite{gong202470} has emerged as a pioneering initiative, specifically focusing on Mandarin stuttering event detection and automatic speech recognition (ASR). Participants involved in the stuttering event detection task are challenged to develop advanced multi-label classification models capable of identifying stuttering instances within short, disfluent speech recordings.

In this paper, we will introduce our submitted stuttering event detection system for Task 1 in the StutteringSpeech Challenge, designated as T031. Notably, annotations for the MSED task are primarily conducted at the audio level, presenting a crucial challenge for existing methods to tackle performance issues stemming from ambiguous audio frames (\textit{confusing frames}). Specifically, when audio contains indistinguishable frames, they are prone to misclassification, which inevitably impacts the network's learning of the representation for stuttering. Ideally, we aim for the network to produce frame embedding features that are more discernible (\textit{easy frames}). We argue that leveraging easy frames within audios to refine confusing frames can facilitate a more accurate distinction between stuttered and fluent speech. We employ a simple yet effective classifier network to model the frame-level likelihood of stuttering events occurring from the backbone network's output. Subsequently, considering the inherent temporal structure of an event (the beginning, climax, and ending) \cite{zheng2022weakly}, we introduce a mining algorithm that samples confusing frames through cascaded contraction and expansion operations on pseudo-boundaries, while easy frames are identified using top-$k$ and bottom-$k$ selection methods.
We then propose a stutter contrast loss to refine the representation of these confusing frames, thereby enhancing discrimination between stuttered and fluent frames. Compared with DisfluentSiam \cite{mohapatra2023ssl1}, we conduct contrastive learning at a finer granularity, eliminating the necessity for data augmentation. Moreover, contrasting different frames within the same audio can provide more comprehensive information for the model.

Our contributions are as follows: (1) We propose a Fine-Grained Contrastive Learning method for MSED, which refines the representation of confusing frames. (2) We model the frame-level likelihood and introduce a mining algorithm to identify the easy and confusing frames within the audio. (3) We extensively evaluate FGCL on the Mandarin benchmark dataset. The results demonstrate the effectiveness of FGCL. Post-contest experimentation with English datasets and SSL features affirms the versatility of our approach.

\section{Baseline System}
\label{sec:format}

The baseline system adopts the Conformer block \cite{gulati2020conformer}, which includes a multi-headed self-attention module, a convolution module, and a feed-forward module. For a given audio $A_n$, its original audio feature input is \( X_n^o \in \mathbb{R}^{T \times D} \), where \( T \) represents the number of frames and \( D \) denotes the dimension of the feature vector, the Conformer block encodes this input to obtain the embedding \( X_n \in \mathbb{R}^{T \times D} \):
\begin{equation}
\label{equ:encode}
X_n = {Conformer}(X_n^o)
\end{equation}
Subsequently, \( X_n \) is processed through an average pooling layer followed by a linear layer and a sigmoid activation function:
\begin{equation}
\label{equ:predict}
y_n = Sigmoid({Linear}({AvgPool}(X_n)))
\end{equation}
\( y_n \in \mathbb{R}^{C} \) represents the probabilities of occurrence for $C$ classes of stuttering events at the audio level. The model is optimized using the MultiLabelSoftMarginLoss function with the ground truth \( y_n^{gt} \):
\begin{equation}
\mathcal{L}_{cls} = {MultiLabelSoftMarginLoss}(y_n, y_n^{gt}) 
\end{equation}
Although Conformer models both local and global dependencies of the audio sequence, it does not fully utilize the inherent temporal structure of stuttering events and may suffer from performance degradation due to confusing frames.

\begin{figure*}[htbp]
	\centering
	\includegraphics[width=0.90\textwidth]{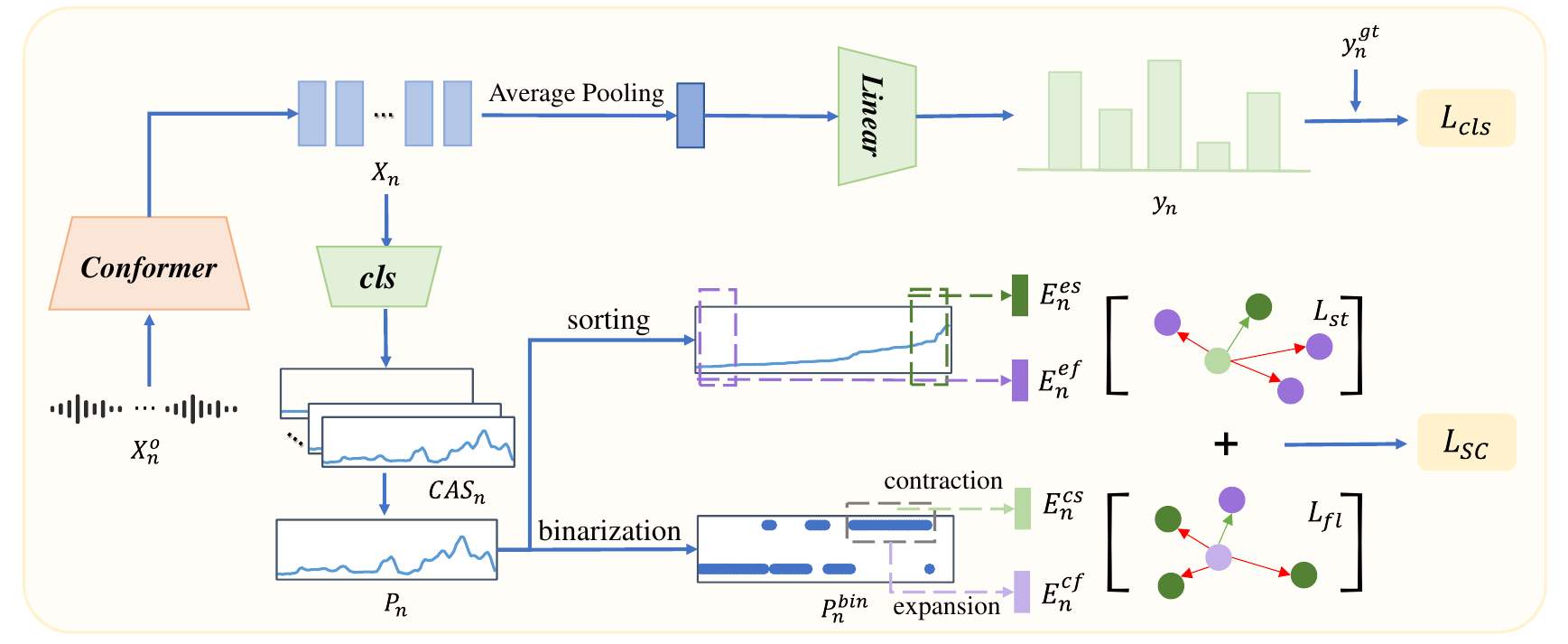}
	\caption{Overview of our FGCL system. Following frame-level likelihood modeling, we mine confusing and easy frames. $\mathcal{L}_{SC}$ aims to refine the confusing embeddings.}
	\label{fig:model}
\end{figure*}

\section{PROPOSED METHOD}

Figure \ref{fig:model} illustrates the schematic diagram of our MSED system. 
As shown in Figure \ref{fig:model}, FGCL provides a plug-and-play module that integrates seamlessly into the existing baseline system. 
Specifically, after obtaining the frame-level embeddings \( X_n \) through the Conformer in the Baseline system, we perform likelihood modeling. This is achieved by using a simple classifier to determine the frame-level likelihood of stuttering events. Subsequently, we employ a cascaded contraction and expansion algorithm to sample confusing frames within the audio. Meanwhile, we use a top-$k$ and bottom-$k$ selection method to sample easy frames within the audio. Finally, we conduct contrastive learning between different frames within the audio to refine the representation of confusing frames.

\subsection{Likelihood modeling}
We model the likelihood of detecting stuttering instances in each frame. Given the diverse representation of various stuttering events, we first utilize a classifier $cls$ to obtain frame-level Class Activation Scores (CAS). 
\begin{equation}
CAS_n = cls(X_n)
\end{equation}
\( cls(\cdot) \) consists of several temporal convolution layers and the ReLU activation function. After that, we sum \( CAS_n \in \mathbb{R}^{T \times C} \) along the channel dimension followed by the sigmoid function to derive the frame-level probabilities of stuttering events:
\begin{equation}
P_n = Sigmoid \left( \sum_{c=1}^{C} CAS_n \right)
\end{equation}
\( P_n\in \mathbb{R}^{T} \) represents the frame-level probabilities of stuttering events.

\subsection{Confusing frames mining}

\begin{figure}[t!]
	\centering
	\includegraphics[width=0.95\columnwidth]{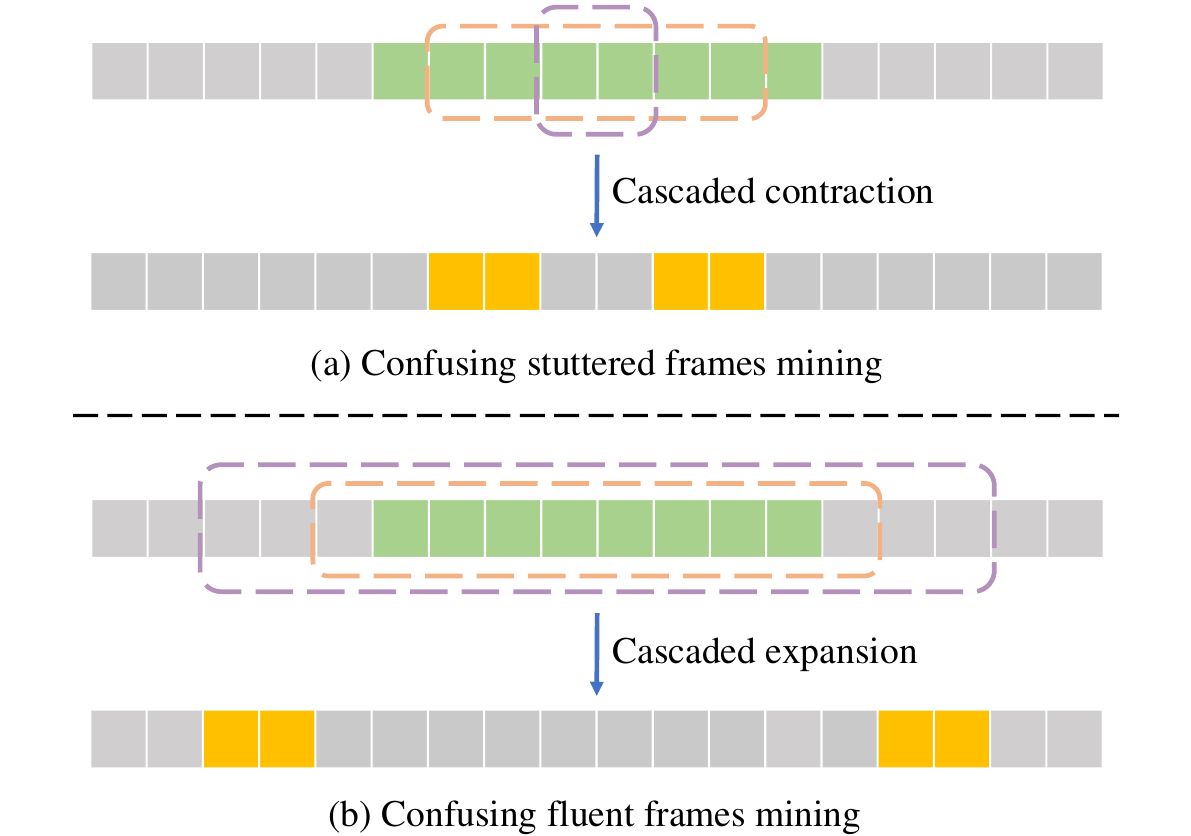}
	\caption{A toy example of mining the confusing frames. The orange dashed box represents the smaller mask ($m=2$), while the purple dashed box represents the larger ($M=4$). We sample confusing frames from the borders of the pseudo-boundaries. }
	\label{fig:mining_algorithm}
\end{figure}

We classify stuttering events into three phases: beginning, climax, and ending, reflecting their inherent temporal progression. The ambiguity in frames near the beginning and ending stems from their transitional characteristics, contrasting with the clarity of frames at the climax. To address the network's reduced accuracy in identifying these ambiguous boundary frames, we firstly need to delineate stuttering event pseudo-boundaries accurately and then sample potentially confusing frames at these edges. Considering the inherent uncertainty in defining pseudo-boundaries, we employ a cascaded algorithm to refine the sampling of frames that might have been incorrectly classified. Initially, we binarize \(P_n\) to localize stuttering event instances.

\begin{equation}
P_n^{bin}=\epsilon({P}_n-\theta)
\end{equation}
where $\epsilon(\cdot)$ denotes the Heaviside step function, and $\theta$ represents the threshold value for the binarization. Subsequently, we mine confusing frames from the borders of the pseudo-boundaries, i.e., the possible stuttering event intervals, in the \( P_n^{bin} \) sequence where the value is 1. As illustrated in Figure \ref{fig:mining_algorithm}, we apply a cascaded contraction with a smaller mask \(m\) and a larger mask \(M\) on the pseudo-boundaries. The difference frame between the two is used to sample the confusing stuttered frames $X_{n}^{st}$. Similarly, we perform a cascaded expansion, where the difference frames are the confusing fluent frames $X_{n}^{fl}$:
\begin{equation}
X_{n}^{st}=(P_{n}^{bin};m)^{-}-(P_{n}^{bin};M)^{-}
\end{equation}
\begin{equation}
X_{n}^{fl}=(P_{n}^{bin};M)^{+}-(P_{n}^{bin};m)^{+}
\end{equation}
$(\cdot;l)^{-}$ and $(\cdot;l)^{+}$ reprensents the contraction and expansion operation using mask $l$ \cite{gil2002efficient}. Next, we sample \(k^c\) frames from the confusing frames to form the confusing stuttered feature $E_{n}^{cs} \in \mathbb{R}^{k^c \times d} $ and the confusing fluent feature $E_{n}^{cf} \in \mathbb{R}^{k^c \times d} $, where \(k^c = \max(1, \left\lfloor \frac{T}{\gamma^{c}} \right\rfloor)\). The parameter \(\gamma^c\) controls the proportion of selected frames.

\subsection{Easy frames mining}
To mine easy stuttered frames and easy fluent frames, we simply utilize the frame-level probabilities of stuttering events \(P_n\). Specifically, we apply a simple top-$k$ and bottom-$k$ selection method on \(P_n\). Frames with the highest probabilities (top-$k$) are sampled as easy stuttered frames, while frames with the lowest probabilities (bottom-$k$) are sampled as easy fluent frames. The easy stuttered feature $E_{n}^{es} \in \mathbb{R}^{k^e \times d} $ and the easy fluent feature $E_{n}^{ef} \in \mathbb{R}^{k^e \times d} $ are defined as follows:
\begin{equation}
E_{n}^{es}=\{X_{n;t}\mid t \in \mathcal{I}_{n}^{st}, \mathcal{I}_{n}^{st}=P_{n}^{Desc}[:k^e]\}
\end{equation}
\begin{equation}
E_{n}^{ef}=\{X_{n;t}\mid t \in \mathcal{I}_{n}^{fl}, \mathcal{I}_{n}^{fl}=P_{n}^{Asc}[:k^e]\}
\end{equation}
\(P_{n}^{Desc}\) and \(P_{n}^{Asc}\) represent the indices of \(P_n\) sorted in descending and ascending order, respectively. The parameter \(k^e\), which controls the proportion of selected frames, is defined as \(\max(1, \left\lfloor \frac{T}{\gamma^{e}} \right\rfloor)\).

\subsection{Stutter contrast loss}
Contrastive learning is a crucial method in representation learning, aiming to learn effective feature representations by contrasting positive and negative pairs of samples. In order to refine the confusing feature representation, we propose a stutter contrast loss. This loss function includes two components: \(\mathcal{L}_{st}\) for refining confusing stuttered feature and \(\mathcal{L}_{fl}\) for refining confusing fluent feature. Given the confusing feature and easy feature, we construct two contrastive pairs. Formally, we denote the query as $\boldsymbol{x}$, the positive sample as $\boldsymbol{x^+}$, and the negative sample as $\boldsymbol{x^-}$. If $\boldsymbol{x} \in E_n^{cs}$ then $\boldsymbol{x^+} \in E_n^{es}$, $\boldsymbol{x^-} \in E_n^{ef}$. \(\mathcal{L}_{st}\) is then calculated as follows:

\begin{equation}
\label{equ:Lst}
\begin{aligned}
\mathcal{L}_{st} = -\log \left[\frac{e^{(E_n^{cs^{T}} \cdot E_n^{es} / \tau)}}
{e^{(E_n^{cs^{T}} \cdot E_n^{es} / \tau)} + \sum_{t=1}^{k^e} e^{(E_n^{cs^{T}} \cdot E_{n,t}^{ef} / \tau)}} \right]
\end{aligned}
\end{equation}
where $\tau$ is the temperature hyperparameter \cite{he2020momentum}, $E_{n,t}^{ef}$ represents the $t$-th negative embedding in easy fluent feature $E_{n}^{ef}$. Like \(\mathcal{L}_{st}\), if $\boldsymbol{x} \in E_n^{cf}$ then $\boldsymbol{x^+} \in E_n^{ef}$, $\boldsymbol{x^-} \in E_n^{es}$. We can compute the contrast loss \(\mathcal{L}_{fl}\) to refine the confusing fluent feature:

\begin{equation}
\label{equ:Lnst}
\begin{aligned}
\mathcal{L}_{fl} = -\log \left[ \frac{e^{(E_n^{cf^{T}} \cdot E_n^{ef} / \tau)}}
{e^{(E_n^{cf^{T}} \cdot E_n^{ef} / \tau)} + \sum_{t=1}^{k^e} e^{(E_n^{cf^{T}} \cdot E_{n,t}^{es} / \tau)}} \right]
\end{aligned}
\end{equation}
The final stutter contrast loss $\mathcal{L}_{SC}$ is obtained by aggregating \(\mathcal{L}_{st}\) and \(\mathcal{L}_{fl}\):
\begin{equation}
\mathcal{L}_{SC}=\mathcal{L}_{st}+\mathcal{L}_{fl}
\end{equation}
\subsection{Network training and inference}
\subsubsection{Training Objective}
Compared to the baseline system, our FGCL incorporates a stutter contrast loss, achieving significant performance improvements in the MSED task. Mathematically, the total loss is the sum of the two losses, expressed as follows:
\begin{equation}
\label{equ:loss}
\mathcal{L} = \mathcal{L}_{cls} + \alpha \mathcal{L}_{SC}
\end{equation}
where $\alpha$ balances the two losses. Since the dataset contains some clips without any stuttering events, we exclude these instances and calculate \(\mathcal{L}_{SC}\) only for the clips containing stuttering events. 
\subsubsection{Inference}
Our primary focus is on utilizing stutter contrast loss to refine the distribution of embeddings in feature space encoded by the backbone network. Therefore, our inference process remains consistent with the baseline system. For a given audio $A_i$, through Eq.(\ref{equ:encode}) and Eq.(\ref{equ:predict}), we can obtain the audio-level categorical probabilities $y_i$. Subsequently, we apply a threshold for each category. If the corresponding value in \(y_i\) exceeds this threshold, we predict the presence of the stuttering event for that category in the audio.

\section{EXPERIMENTS}
\subsection{Experiment settings}
\subsubsection{Dataset}

We utilize the data provided by the StutteringSpeech Challenge \cite{gong202470} to complete Task 1: stuttering event detection. This dataset comprises approximately 50 hours of audio segments and their corresponding labels. We strictly adhere to the challenge rules by using the officially designated training set for training and the official test set for testing.

Beyond the challenge, we also evaluate our method on two English datasets: SEP-28k \cite{lea2021sep} and FluencyBank \cite{ratner2018fluency}, to verify the robustness and effectiveness of our method. SEP-28k contains 28,177 clips extracted from publicly available podcasts. FluencyBank includes recordings of 32 adults who stutter, and we utilize the 4,144 clips annotated by \cite{lea2021sep}. For the two English datasets, we divided them into training, validation, and test sets in a ratio of 8:1:1.

\subsubsection{Implementation details}
Following the baseline system, all audio data is resampled to 16000Hz. We extract 80-dimensional fbank features with a frame length of 25ms and a frame shift of 10ms. Additionally, spectral masking is applied within a range of up to 50 frames and up to 10 frequency bins, enhancing the model's robustness and generalization capability. The backbone network comprises three Conformer blocks. The classifier is constructed with two temporal convolution layers, each having an output dimension of 5, with kernel sizes of 9 and 7, respectively. We set \(m=3\) and \(M=6\) for confusing frame mining. We set \(\gamma^c=10\) and \(\gamma^e=20\) for frame selection (\(\gamma^e=10\) in our official result). We set \(\tau=0.07\) in Eq.(\ref{equ:Lst}) and Eq.(\ref{equ:Lnst}). We set \(\alpha=0.05\) in Eq.(\ref{equ:loss}). For evaluation, we employ the standard protocol by reporting the F1 score, consistent with the baseline system's evaluation method. All experiments are run on an NVIDIA Tesla A30 GPU.

\subsection{Experimental results for StutteringSpeech Challenge}
The F1 score is the evaluation metric for the challenge. The test dataset includes five types of stuttering events: sound prolongation (/p), sound repetition (/r), word repetition (/wr), block (/b), and interjection (/i). We calculate the F1 score for each of these five types individually and compute the average. Table \ref{table:comapre baseline} presents the official results of the baseline and our FGCL, as well as the results of FGCL after parameter adjustments. The results in Table \ref{table:comapre baseline} demonstrate that our method is more accurate in detecting all types of events, resulting in higher F1 scores. Specifically, the official results show that the average F1 score of our method is about 3.0\% higher than the baseline. After parameter adjustments, FGCL improves the F1 score by over 5.1\% compared to the baseline, representing a significant improvement.
\begin{table}
	\caption{Official results in the StutteringSpeech Challenge in SLT2024. * denotes the re-implementation with parameter adjustments.}
	\label{table:comapre baseline}
	\centering
 \resizebox{\linewidth}{!}{
	\begin{tabular}{l|*{5}{c}|c}
		\hline 
		\multirow{2}{*}{Method} & \multicolumn{5}{c|}{F1 score} & \multirow{2}{*}{F1 (Avg)} \\ 
		& /p &/b & /r & /wr& /i & \\
		\hline
  Baseline& 65.12&24.30&41.86&61.85&74.87&53.60\\
		 FGCL& 65.89&24.78&50.15&\bf{64.85}&77.06&56.55	\\
		FGCL\textsuperscript{*}& \bf{66.10}&\bf{33.64}&\bf{53.65}&62.99&\bf{77.15}& \bf{58.70}\\
		\hline
	\end{tabular}
 }
\end{table}
\subsection{Ablation studies}
\subsubsection{The effect of each component of stutter contrast loss}
Our proposed stutter contrast loss $\mathcal{L}_{SC}$ consists of two components: $\mathcal{L}_{st}$ and $\mathcal{L}_{fl}$. To validate the effectiveness of each component, we conduct an ablation study by individually removing each component. The results in Table \ref{table:ablation component} demonstrate that each individual loss term significantly improves performance. Furthermore, the combination of both loss terms yields better results, highlighting the synergistic effect of the two components.
\begin{table}
	\caption{Ablation study of components in stutter contrast loss.}
	\label{table:ablation component}
	\centering
 \resizebox{\linewidth}{!}{
	\begin{tabular}{l|*{5}{c}|l}
		\hline 
		\multirow{2}{*}{Method} & \multicolumn{5}{c|}{F1 score} & \multirow{2}{*}{F1 (Avg)} \\ 
		& /p &/b & /r & /wr& /i & \\
		\hline
  Baseline& 65.12&24.30&41.86&61.85&74.87&53.60\\
		\hline 
		 $+\mathcal{L}_{fl}$& 67.46&27.40&46.41&64.71&77.02&56.60{\textcolor{blue}{\scriptsize +3.00}}	\\
		 $+\mathcal{L}_{st}$& 66.50&31.75&51.48&64.25&77.09&58.21{\textcolor{blue}{\scriptsize +4.61}}	\\
		$+\mathcal{L}_{fl}+\mathcal{L}_{st}$& {66.10}&{33.64}&{53.65}&62.99&{77.15}& 58.70{\textcolor{blue}{\scriptsize +5.10}}\\
		\hline
	\end{tabular}
 }
\end{table}

\subsubsection{Analysis of the confusing frames mining algorithm}

We validate the effectiveness of our cascaded contraction and expansion strategy through two variants. Note that when $m=1$, the smaller mask will not affect the sequence. We evaluate two variants: one keeping $M$ the same as in Line \#3, and the other maintaining the same mining length. From the results in Table \ref{table:ablation cascaded}, it is evident that our adoption of a cascaded strategy during contraction and expansion processes is effective. Due to noise from pseudo-boundaries, frames closely bordering the boundaries may not be accurate. Our cascaded strategy effectively filters these edge frames, improving the accuracy of confusing frame selection.


\begin{table}
	\caption{Ablation study of the cascaded mining algorithm.}
	\label{table:ablation cascaded}
	\centering
    \resizebox{\linewidth}{!}{
	\begin{tabular}{c|cc|*{5}{c}|c}
		\hline 
		\multirow{2}{*}{\#} &\multirow{2}{*}{$m$} & \multirow{2}{*}{$M$} & \multicolumn{5}{c|}{F1 score} & \multirow{2}{*}{F1 (Avg)} \\ 
		&& & /p & /b & /r & /wr & /i & \\
		\hline
		 1 &1 & 6 & 66.91	&26.4	&48.57	&61.97	&78.59 & 56.49\\
		 2 &1 &4 & 67.69&	28.45&	47.53&	64.13&	77.11 & 56.98 \\
		 3 &3 & 6 & 66.10 & 33.64 & 53.65 & 62.99 & 77.15 & 58.70\\
		\hline
	\end{tabular}
    }
\end{table}

\subsection{Generalization of the proposed FGCL}
To further verify the generalization of our proposed FGCL, 
We utilize the pre-trained model HuBERT \cite{hsu2021hubert} to extract embedding features, replacing the original audio feature input. As shown in Table \ref{table:ablation English}, even though using pre-trained features improves the detection accuracy of stuttering events, FGCL can further boost its performance through frame-level stutter contrast loss. Then we evaluate both the Baseline system and FGCL on two benchmark English stuttering datasets. The results in Table \ref{table:ablation English} demonstrate that our proposed FGCL can consistently improve their performance. This indicates the strong generalization capability of FGCL across different feature types and languages, and also validates the effectiveness of our algorithms for mining confusing and easy frames.

\begin{table}
	\caption{Evaluation of the generalization of FGCL. * denotes using features extracted with HuBERT.}
	\label{table:ablation English}
	\centering
    \resizebox{\linewidth}{!}{
	\begin{tabular}{c|c|*{5}{c}|c}
		\hline 
		\multirow{2}{*}{Dataset} & \multirow{2}{*}{Method} & \multicolumn{5}{c|}{F1 score} & \multirow{2}{*}{F1 (Avg)} \\ 
		&& /p & /b & /r & /wr & /i & \\
		\hline
  \multirow{2}{*}{Mandarin\textsuperscript{*}}
		 &Baseline & 67.89&39.55&51.9&62.7&82.87& 60.98\\
		& FGCL & 69.28&40.55&57.42&65.12&83.18 & 63.11 \\
   		\hline
    \multirow{2}{*}{SEP-28k}
		 &Baseline & 61.39& 61.74& 53.02& 39.52& 68.55& 56.84\\
		& FGCL & 61.96& 64.98& 55.75& 41.74& 67.93 & 58.74 \\
   		\hline
    \multirow{2}{*}{FluencyBank}
		& Baseline &50.98&55.09&51.12&38.89&68.69& 52.96\\
		& FGCL & 51.16&57.75&54.0&45.79&72.13 & 56.17\\
		\hline
	\end{tabular}
    }
\end{table}

\section{Conclusion}
\label{sec:foot}
In this paper, we present a Mandarin Stuttering Event Detection system for the StutteringSpeech Challenge in SLT2024, denoted as T031. To explore the finer granularity of audio, we propose a novel Fine-Grained Contrastive Learning framework for MSED. Our approach utilizes frame-level fine-grained contrasting to enhance the accuracy of stuttering event detection. Initially, we model the likelihood of stuttering events at the frame level. Subsequently, we introduce a mining algorithm to identify confusing and easy frames within the audio. Lastly, we propose a stutter contrast loss to refine the representation of confusing frames, thereby improving the distinction between stuttered and fluent speech. Extensive experiments conducted on benchmark datasets in both English and Mandarin validate the effectiveness of our approach.

\bibliographystyle{IEEEbib}
\bibliography{refs}

\end{document}